\documentstyle[12pt,aps]{revtex}

\def\be{\begin{equation}}
\def\ee{\end{equation}}

\newcounter{fig}

\begin{document}

\title{\LARGE \bf Generalized model for dynamic percolation}

\author{\Large O.B\'enichou$^{1,3}$, J.Klafter$^2$, M.Moreau$^3$ and  G.Oshanin$^3$}

\address{$^1$ Laboratoire de Physique Th\'eorique et Mod\`eles Statistiques, \\
Universit\'e Paris-Sud, 91405 Orsay Cedex, France
}

\address{$^2$ School of Chemistry,\\ Tel Aviv University,  
Tel Aviv 69978, Israel
}

\address{$^3$ Laboratoire de Physique Th\'eorique des Liquides,\\
 Universit\'e Paris 6, 
4, Place Jussieu, 75252 Paris, France
}
\begin{tightenlines}
\address{\rm (Received: )}
\address{\mbox{ }}
\address{\parbox{16cm}{\rm \mbox{ }\mbox{ }
We study the dynamics of a carrier, which performs
a biased motion under the influence of an external field ${\vec E}$,  in an
environment which is modeled by dynamic percolation and  
created by 
hard-core particles. The particles 
move randomly  
on a simple cubic lattice, constrained by hard-core exclusion,
and they spontaneously annihilate and re-appear  at some prescribed
rates.
Using  decoupling 
of the third-order 
correlation functions 
into the product of the
pairwise carrier-particle correlations  
we determine the
density profiles of the "environment" particles, 
as seen from the stationary moving carrier,  and
calculate
its terminal velocity, $V_{c}$,  as the function of the applied 
field and other system parameters.  
We find that for sufficiently small driving forces 
the force exerted  on the carrier by the "environment"
 particles shows a viscous-like behavior. 
An analog  Stokes formula for 
such dynamic percolative environments and  
 the corresponding friction coefficient are derived.   We show  that
 the density profile of the environment particles is
strongly inhomogeneous:
In front of the stationary
 moving carrier the density is higher than the average density, $\rho_s$, and approaches the
average
value as an exponential function of the distance from the carrier.
Past the carrier the local density is lower than $\rho_s$ and
 the relaxation towards  $\rho_s$ may proceed differently depending on whether the particles
number is or is not explicitly conserved. 
}}
\address{\mbox{ }} 
\address{\parbox{16cm}{\rm PACS numbers:  05.40.+j, 
05.60.+w, 02.50.+s, 05.70.Ln, 47.40. Nm}}
\maketitle

\makeatletter
\global\@specialpagefalse

\makeatother
\end{tightenlines}

\section{Introduction.}

The percolation concept has turned out very useful for understanding
transport
and conduction processes in a wide range of disordered media, 
as exemplified by  
ionic conduction 
in polymeric, amorphous or glassy ceramic  electrolytes, 
diffusion in biological tissues and
permeability of disordered membranes \cite{1,2,3}.

Most of the situations discussed in Refs.\cite{1,2,3}
pertain, however, to  systems 
with "frozen" disorder; that is, 
the random
environment in which a given 
transport process takes place does not
change in time. This is certainly 
the case in many instances, but it is not true 
in general.
As a matter of fact, there are many
experimental systems in which the static percolation 
picture does not apply
since the structure of the host material 
 undergoes
essential structural reorganizations on a 
time scale comparable to that at which the
transport itself occurs. A few
stray examples of such systems include certain biomembranes \cite{4}, 
solid protonic conductors \cite{5},
oil-continuous microemulsions \cite{6,7,8,9} and polymer
electrolytes \cite{10,11,12}. 

More specifically, ionic transport across a biomembrane, such as,
 e.g. gramicidin-$A$,
 occurs by the motion of ions through
molecular channels along which they encounter potential barriers
 that fluctuate in time. The fluctuations of
potential barriers may hinder significantly the transport and
 constitute
an important transport-controlling  factor   \cite{4}. 
In the case of protonic conduction by
the Grotthus mechanism \cite{5}, site-to-site motion of 
carriers occurs only between those neighboring
$H_2O$ or $NH_3$ groups that have a favorable
 relative orientation; thermally activated rotation of
these groups is the structural host-reorganization process
 interacting with the carrier motion. Similarly,  within
 oil-continuous microemulsions,
 the charge transport
proceeds by charge being transfered from one water globule
 to another, as globules approach each other in
their Brownian motion \cite{6,7,8,9}. Lastly, in 
polymer electrolytes, such as, e.g. polyethylene
oxide complexed nonstoichiometrically with the ionic salt
$NaSCN$, the $Na^+$ ions are largely
tetrahedrally coordinated by polyether oxygenes, but
 at the same time that $Na^+$ ions hop from
one fourfold coordination site to another, the oxygens
 themselves, along with the polymeric
backbone, undergo large-amplitude wagging and even
 diffusive motion \cite{10,11,12}.

Clearly, all the above mentioned examples involve two characteristic
 time scales, one which describes the typical
time $\tau$ between two successive hops of the carrier, while the other
is associated with a typical renewal time
 $\tau^*$ of the environment itself; namely,  the time needed 
 for the host medium to re-organize itself
 and thereby  provide a new
set of available pathways for transport. 
 Consequently, the static percolation picture
applies only  
when the  characteristic time $\tau^*$  gets
infinitely large.
For a finite $\tau^*$  $dynamic \; percolation$ has to be considered, and 
one encounters  quite a different behavior when
compared to the random environments with quenched
disorder. As a result,  one observes  Ohmic-type 
or  Stokes-type linear velocity-force relation 
for the carrier's  terminal
velocities as a function of the applied field, in contrast
 to the threshold behavior predicted by the static
percolation theory.  
The prefactor in the
 linear velocity-force  relation may depend, however, in a non-trivial way 
on the system's parameters and this  dependence consitutes  the
main challenge for the theoretical analysis here.
On the other hand, we note that in the above
mentioned examples of the dynamic percolative environments quite
different physical processes are responsible for the time evolution of the host medium.
Consequently, one expects that the prefactor in the Stokes-type 
velocity-force
relation should also be dependent 
 on the precise mechanism which underlies the temporal
re-organization of the environment.

Theoretical modelling of dynamic percolative environments has followed 
several avenues, which differ mostly in how  the time evolution 
 of the disorder is constrained; Namely, is it constrained  (a) by  conservation laws or (b)  by
spatial and temporal correlations
in the renewal events? Early models 
 of dynamic percolation \cite{drugerA,drugerB} described the
random environment
within the framework of a standard bond-percolation
model, in which the strength of each bond  fluctuates 
in time between zero and a
finite value.  The dynamics of the host medium in these models 
 \cite{drugerA,drugerB} was accounted for  by a series of instantaneous
renewal events. These events were assumed to occur at random times,
chosen from  a renewal time
 distribution. In the renewal process the
positions of all unblocked bonds are being reassigned,
 such that after each renewal 
event a carrier sees a newly defined
network.  This approach is thus characterized by
a $global \; dynamical \; disorder$ without global conservation laws and
correlations, 
 since the entire
set of random hopping rates is simultaneously renewed independently of the
previous history. 
Another model characterized by 
a $local \; dynamical \; disorder$ has been proposed in Refs.\cite{zwanzig}
 and
\cite{sahimi}, and subsequently generalized to
 the non-Markovian case in Ref.\cite{chatterjee}. 
This model appears to be 
similar to the previous one,
except that here the hopping rates at different
 sites fluctuate $independently$ of each other. 
That is, individual bonds, rather than the whole 
lattice change in the renewal events.
To describe the dynamical behavior in the $local \; dynamical \; disorder$ 
case,   a dynamical
 mean-field theory has been proposed \cite{zwanzig,sahimi},
based on the effective medium
approximation introduced for the
 analysis of random walks on lattices with static
disorder \cite{odagaki}, and has been generalized  to include the
 possibility of multistate transformations of the
$dynamically$ random medium \cite{granekB}. More recently, 
 several exactly
solvable one-dimensional models with $global$ and
 $local$  dynamical disorder have been discussed \cite{garcia}.

In the second  approach, which emerged within the context of the ionic conductivity in
superionic solids, 
the dynamical 
percolative environment has been considered as a multicomponent mixture 
of mobile
species in which one or several neutral components  block
 the carrier component   \cite{hilferA}.  In particular, such a situation can be observed in
 a superionic
conductor $\beta''$-alumina, doped with two different ionic species (e.g. $Na^+$ and $Ba^{2+}$),
where small $Na^+$ ions are rather mobile, while
the larger $Ba^{2+}$ ions move essentially slower and temporarily 
block the $Na^+$ ions. 
Contrary to the previous line of thought, the
dynamics of such a percolative environment has essential correlations, generated by hard-core
exclusion interactions  between the species involved, and moreover, it obeys the conservation law
- the total number of the particles involved is conserved. In Ref.\cite{hilferA}, the
frequency-dependent ionic conductivity of the light species has been analysed combining a
continuous time random walk approach for the dynamical problem with an effective medium
approximation describing the frozen environment of slow species.
Next, as an explanation of the sharp increase of 
electrical conductivity transition in water-in-oil microemulsions when the volume fraction of
water is increased towards a certain threshold value, 
in Refs.\cite{7} and \cite{8} it has been proposed that the charge carriers are not trapped in the
finite water clusters, but rather a charge on a water globule can propagate by either hopping to a
neighboring globule, when they approach each other, or via the diffusion of the host globule
itself. This picture has been interpreted in terms of a model similar to that employed in
Ref.\cite{hilferA}, with the only difference being that here the "blockers" of Ref.\cite{hilferA} 
play the role of the
transient charge carriers. In the model of Refs.\cite{7} and \cite{8}, in which 
the host dynamics is influenced by
spatial correlations and conservation of the number of the
water globules involved,  the conductivity depends hence, on the rate of cluster rearrangement.     
Lastly, a similar problem of a carrier diffusion in an environment created by mobile hard-core
 lattice-gas particles has
been analysed in Ref.\cite{granekS} by using the developed dynamic bond
 percolation theory  of Refs.\cite{drugerA} and
\cite{drugerB}.

In this paper we propose a generalized model of dynamic percolation which shares common features
with both bond-fluctuating models of Refs.\cite{drugerA,drugerB,zwanzig,sahimi,chatterjee,granekB,garcia} 
as well as models involving mobile blockers of
Refs.\cite{hilferA,granekS}.
The system we consider 
consists of
a host lattice, which here is a regular
cubic lattice whose sites 
support at most a single occupancy, 
hard-core "environment" particles,    
and a single hard-core carrier
particle. 
The "environment" particles move 
on the lattice
by performing a random
hopping 
between the neighboring lattice sites,
which is constrained by the 
hard-core interactions, 
and may disappear from and re-appear (renewal processes) on the empty sites of the lattice
with  some prescribed rates\footnote{We hasten to remark that diffusive processes, of
course, also result in a certain renewal of the environment; 
diffusive processes, as compared to the spontaneous creation and annihilation 
of particles, have however 
completelely different underlying physics and influence in a
completely different fashion the evolution of the system, as we proceed to show.
Following the terminology of Refs.\cite{drugerA,drugerB,zwanzig,sahimi,chatterjee,granekB,garcia}, 
we thus choose here to distinguish  
between diffusive and creation/annihilation processes, referring to the latter as the renewal ones.}.
In turn, the
carrier particle  is  always present on the lattice, i.e., 
it can not disappear spontaneously, and is
subject to
a constant external force ${\vec E}$. Hence, 
the carrier performs
a biased random walk,  
which is
constrained by the  hard-core
interactions with the
"environment" particles, 
and  probes 
the response of the percolative environment 
to the internal perturbancy or, in other words,
the frictional properties of such a dynamical environment. 

An important aspect  of our model, which makes it different to the previously proposed models of dynamic
percolation, is that we include the hard-core 
interaction between "environment" particles and the 
carrier molecule, such that the latter may influence the dynamics of the 
environment. This results, as we proceed to show, in the emergence of complicated density profiles of the
"environment" particles around the carrier. These profiles, as well as the terminal velocity  $V_{c}$ 
of the carrier, are determined here explicitly, 
in terms of an approximate
approach of Ref.\cite{burlatsky}, which is based on
the decoupling of 
the carrier-particle-particle
 correlation functions into the product of pair-wise 
correlations. 
We show that the 
"environment" particles
tend to accumulate in front of the driven carrier 
creating a sort of a "traffic jam", which impedes
 its motion. Thus the density profiles  around the carrier are
highly asymmetric: the local density of the "environment"
 particles in front of the carrier is higher than the 
average and approaches the average value as an exponential
 function of the distance from the carrier. The characteristic
length and the amplitude of the density relaxation function
are calculated explicitly.
On the other hand, past the carrier 
the local density is lower than the average: We show that depending on
 the condition whether the number of particles in the percolative environment 
is explicitly conserved or not, the local density past the carrier
 may tend to the average value either as an exponential or even as an
 $\em algebraic$ function of the distance, revealing in the latter case
especially strong memory effects and strong 
correlations between the particle distribution in the
 environment and the carrier position. Further on, we find
 that the terminal velocity of the carrier particle depends
 explicitly on the excess density in the "jammed" region in
 front of the carrier, as well as on the "environment" particles density past the carrier. Both, in turn, are
dependent on the
 magnitude of the velocity, 
as well as on the rate of the renewal processes and the
 rate at which the "environment" particles can diffuse away from the carrier.
 The interplay between the jamming effect of the environment, produced by the carrier 
particle, and the rate of its
homogenization due to diffusive smoothening  and  renewal processes, manifests itself
 as a medium-induced frictional force exerted on the carrier, whose 
magnitude depends on the carrier velocity. 
As a consequence of such a non-linear coupling,    
in the general case, (i.e. for arbitrary rates of the renewal and diffusive 
processes), 
$V_{c}$ can be found only implicitly, 
as the solution of a non-linear 
equation relating $V_{c}$ to the system parameters. 
This equation 
simplifies considerably   
in the limit of small applied external fields ${\vec E}$ and
we find that the force-velocity relation to the field becomes linear.
This implies that the frictional force exerted on the carrier
particle by the 
environment is $\em viscous$. This linear force-velocity relation can be therefore
 interpreted as the analog of the Stokes
formula for the dynamic percolative environment under study; in this case, 
the carrier velocity is calculated explicitly as well as the corresponding friction coefficient.
In turn, this enable us to estimate the self-diffusion coefficient of the carrier in
 absence of external field; we
show that when only diffusive re-arrangement of the percolative  environment is 
allowed, while the renewal processes
are suppressed, the
general expression for the diffusion coefficient  reduces to the one obtained 
previously in Refs.\cite{nakazato} and
\cite{elliott}. We note  that the result of Refs.\cite{nakazato} and
\cite{elliott} is known to serve as a very good approximation for the
 self-diffusion coefficient in hard-core
lattice-gases \cite{kehr}.  

We finally remark,
 that a qualitatively similar physical effect was predicted recently
 for a different model system involving a charged particle moving 
at a constant speed  a small distance above the surface of an incompressible,
 infinitely deep liquid. It has been shown in Refs.\cite{elie1,elie2}, 
that the interactions between the moving particle and the fluid molecules
  induce an effective frictional force exerted on the particle, producing
 a local distortion of the liquid interface, - a bump, which travels 
together with the particle and increases effectively its mass.  The mass of the bump, 
which is
analogous to the jammed region appearing in our model, depends itself
on the particle's velocity resulting
in a non-linear coupling  between the medium-induced
 frictional force exerted on the particle and its
velocity \cite{elie1,elie2}.

The paper is structured as follows: In Section
 II we formulate
the model and introduce
basic notations.  In Section III we write down the dynamical
equations which govern the time evolution of the "environment"
particles and of the carrier. Section IV is devoted to the
analytical solution of these
evolution equations in the limit $t \to \infty$ ; here we also
 present some 
general results on the shape of the density profiles around stationary moving carrier and  
on the carrier terminal velocity,
which is given implicitly, as the solution of a
transcendental equation defining the general force-velocity relation
 for the dynamic percolative environment under
study. In Section V
we derive explicit asymptotic results for the carrier terminal velocity 
in the limit of small applied external fields ${\vec E}$ and obtain the analog of the Stokes formula for such
a percolative
environment; as well, we present here explicit results for the friction coefficient of the host medium and
for the self-diffusion coefficient of the carrier in the absence of external field. 
Asymptotic behavior of
the density profiles of the "environment" particles around the carrier is discussed in Section VI. 
Finally, we conclude in Section VII with a brief summary
and discussion of our results.

\section{The model.}

The model for dynamic percolation we study here consists of a three-dimensional
simple cubic lattice of spacing $\sigma$, the sites of which are partially occupied by
identical hard-core "environment" 
particles and a single, hard-core, carrier particle (see Fig.1). 
For both types of particles
the hard-core interactions prevent multiple occupancy of
the lattice sites; 
 that is, no two "environment" particles or the "carrier" and an "environment" particle
 can occupy simultaneously
the same site, and particles can not pass through each other. 

The occupation of the
lattice sites by the "environment" particles
is characterized by the time-dependent occupation variable 
$\eta({\vec r})$, ${\vec r}$ being the lattice-vector of the site in question. This
variable 
assumes two values:
\begin{equation}
\eta({\vec r}) = \left\{\begin{array}{ll}
1,     \mbox{ if the site ${\vec r}$ is occupied} \nonumber\\
0,     \mbox{ if the site ${\vec r}$  is empty}
\end{array}
\right.
\end{equation} 
Next, we assume the following  dynamics of the "environment" 
particles:
The particles 
can
spontaneously disappear from the lattice, and may re-appear at random
positions and random time moments, which is reminiscent of the host medium dynamics stipulated 
in Refs.\cite{drugerA,drugerB,zwanzig,sahimi,chatterjee,granekB,garcia}. We refer to 
these two processes generally as 
renewal processes.
In addition, the environment particles move randomly within the lattice by performing
nearest-neighbor random walks constrained by the hard-core interactions, which is the main feature 
of the approach in Refs.\cite{hilferA,granekS}. 
We stipulate that any of the "environment" particles waits a time $\delta \tau$, which has an exponential
probability distribution with a mean $\tau^*$, and then 
chooses from a few possibilities: 
(a) disappearing from the lattice at rate $g$, which is realized instantaneously,
 or (b) attempting to hop, at rate $l/6$, onto one of $6$ neighboring sites. 
The hop is actually fulfilled if the
target site is not occupied at this time moment by any other particle; otherwise, the
particle attempting to hop remains at its initial position, and (c)  particles may
re-appear on any $\em vacant$ lattice site with rate $f$.

Note that, 
for simplicity, we  assumed  that the characteristic diffusion time and
the renewal times of the "environment" particles
are equal to each other. These times, i.e. $\tau_{dif}$, mean creation time $\tau_{cr}$
and mean annihilation time $\tau_{an}$ may, however, be
different, and can be restored in our final results by a mere replacement
$l \to l \tau^*/\tau_{dif}$, $f \to f \tau^*/\tau_{cr}$ and $g \to g \tau^*/\tau_{an}$.

Note also that the number of particles is not explicitly conserved in such a dynamical
model of the environment, which happens because of the presence of 
 the renewal processes; the particles diffusion, on contrary, conserves the
particles number. However, 
in the absence of attractive particle-particle interactions and external
perturbances, the particles distribution on the lattice is uniform and the average occupation
$\rho(t) = \overline{\eta({\vec r})}$ of the lattice tends, as $t \to \infty$, 
to a constant value, $\rho_s = f/(f + g)$. This relation
 can be thought of as the Langmuir adsorption
isotherm \cite{langmuir}. 

Hence, the limit $\tau_{dif} \to \infty$ (or, $l \to 0$) corresponds to the ordinary site 
percolation model with immobile blocked sites. The limit $f,g \to 0$, ($\tau_{cr}, 
\tau_{an} \to \infty$), while keeping 
the ratio $f/g$  fixed,
$f/g = \rho_s/(1-\rho_s)$, corresponds to the
usual hard-core lattice-gas with the  number of particles conserved.

At time $t = 0$ we introduce at the  origin of the lattice an extra particle, the carrier,
whose role is to probe the response of the  environment modeled by dynamic percolation to an external
perturbance. We stipulate that  only the carrier out of all participating particles 
 can not disappear from the system, and moreover, 
its motion
is biased by some external constant
force. As a physical realization, we envisage that the carrier is charged, while all other particles are neutral,
 and the system is exposed to constant external electric
field ${\vec E}$. The dynamics of the carrier particle is defined as follows: 
We suppose  that the waiting time between successive jumps of the carrier has also an exponential
distribution with a mean value   
$\tau$, which may  in general be different from the corresponding waiting time of the environment particles.
Attempting to hop, the carrier first chooses a hop direction with  probabilities
\begin{equation}
\label{trans}
p_\mu=exp\Big[\frac{\beta}{2}\Big({\vec E } \cdot {\vec e}_{
\mu}\Big)\Big]/\sum_{\nu}exp\Big[\frac{\beta}{2}\Big({\vec E} \cdot {\vec e}_{\nu}\Big)\Big],
\end{equation}
where $\beta$ is the reciprocal temperature,   
 ${\vec e}_{\nu}$ (or ${\vec e}_{\mu}$) stand for six unit lattice vectors, 
 $\nu,\mu = \{\pm1,\pm2,\pm 3\}$,  
connecting the carrier position with $6$ neighboring lattice sites, and 
$({\vec E} \cdot {\vec e}_{\nu})$ denotes the scalar product.
 We adopt the
convention that $\pm 1$ corresponds to
 $\pm X$, $ \pm 2 $ corresponds to $\pm Y$ while $\pm 3$ stands for $\pm Z$.
The jump is actually fulfilled
when the target lattice site is vacant. Otherwise, as mentioned the carrier remains at its position.
For simplicity we  assume in what follows that the external field is
oriented along the $X$-axis in the positive direction, such that ${\vec E} = (E,0,0)$. Note also that for the choice of the transition
probabilities as in Eq.(\ref{trans}), the detailed balance is naturally preserved.

\section{Evolution Equations}
  
Let 
$P({\vec R_{c}},\eta;t)$ denote the joint probability that at time moment $t$ the
carrier occupies position 
${\vec R_{c}}$ and all "environment" particles are in configuration 
$\eta\equiv\{\eta({\vec r})\}$. Next, let 
$\eta^{{\vec r},\mu}$ denote particles' configuration obtained from 
$\eta$ by exchanging the occupation variables of the sites 
${\vec r}$ and ${\vec r}+{\vec e}_{\vec \mu}$, 
i.e. $\eta({\vec r})\leftrightarrow \eta({\vec r}+{\vec e}_{\vec \mu})$, and
$\hat{\eta}^{{\vec r}}$ be the configuration obtained from
$\eta$ by changing the occupation of the site ${\vec r}$ 
as $\eta({\vec r})\leftrightarrow1-\eta({\vec r})$. Clearly, the first type of process
appears due to random hops of the "environment" particles, while the second one stems
from the renewal processes, i.e. random creation and annihilation of the "environment" particles. Then, summing up all possible events which
can result in the configuration 
 $({\vec R_{c}},\eta)$ or change this configuration for any other, we find that the
temporal evolution of the system under study is governed by the following master
equation:
\begin{eqnarray}
&&\partial_tP({\vec R_{c}},\eta;t)=
\frac{l}{6\tau^*}\sum_{\mu}\;\sum_{{\vec r}\neq{\vec R_{c}}-{\vec e}_{\vec \mu},{\vec R_{c}}} 
 \; \left\{ P({\vec R_{c}},\eta^{{\vec r},\mu};t)-P({\vec R_{c}},\eta;t)\right\} +\nonumber\\
&+&\frac{1}{\tau}\sum_{\mu}p_\mu\left\{\left(1-\eta({\vec R_{c}})\right)P({\vec R_{c}}-{\vec e}_{\vec \mu},\eta;t)
-\left(1-\eta({\vec R_{c}}+{\vec e}_{\vec \mu})\right)P({\vec R_{c}},\eta;t)\right\}+\nonumber\\
&+&\frac{g}{\tau^*}\sum_{{\vec r}\neq {\vec R_{c}}} 
\;\left\{\left(1-\eta({\vec r})\right)P({\vec R_{c}},\hat{\eta}^{{\vec r}};t)-\eta({\vec r})P({\vec R_{c}},\eta;t)\right\}+\nonumber\\
&+&\frac{f}{\tau^*}\sum_{{\vec r}\neq{\vec R_{c}}} 
\;\left\{\eta({\vec r})P({\vec R_{c}},\hat{\eta}^{{\vec r}};t)
-\left(1-\eta({\vec r})\right)P({\vec R_{c}},\eta;t)\right\}.
\label{eqmaitresse}
\end{eqnarray}
Note that the terms in the first (resp. second) line of Eq.(\ref{eqmaitresse}) describe random hopping 
motion of the
"environment" particles (resp. biased motion of the carrier) in terms of the Kawasaki-type
particle-vacancy exchanges, while the terms in the third and the fourth lines account for the Glauber-type
decay and creation of the "environment" particles. 
  
\subsection{Mean velocity of the carrier and correlation functions.}

From Eq.(\ref{eqmaitresse}) we can readily compute the  velocity of the carrier. 
Multiplying both sides of Eq.(\ref{eqmaitresse}) by $({\vec R_{c}} \cdot {\vec e_1})$ and summing over all possible
configurations  $({\vec R_{c}},\eta)$ we find that the carrier's mean velocity $V_{c}(t)$, defined as
\begin{equation}
V_{c}(t)\equiv\frac{{\rm d}}{{\rm dt}}\Big(\overline{{\vec R_{c}} \cdot {\vec e_1}}\Big),
\end{equation}
 obeys:
\begin{equation}
V_{c}(t)=\frac{\sigma}{\tau}\left\{p_1(1-k({\vec e_1};t))-p_{-1}(1-k({\vec e_{-1}};t))\right\},
\label{vitesse}
\end{equation}
where $k({\vec \lambda};t)$ stands for the carrier-"environment" particles pair correlation function
\begin{equation}
k({\vec \lambda};t)\equiv\sum_{{\vec R_{c}},\eta}\eta({\vec R_{c}}+{\vec \lambda})
P({\vec R_{c}},\eta;t).
\label{defk}
\end{equation}
In other words,  $k({\vec \lambda};t)$ can be thought of as
 the density distribution of the "environment"
particles as seen from the carrier which moves with velocity $V_{c}(t)$.

Hence, 
$V_{c}(t)$ depends explicitly on the local density of the "environment"
 particles in the immediate vicinity of
the carrier.  Note that if the "environment" is perfectly homogeneous, 
i.e. if for any ${\vec \lambda}$
the density profile is 
constant, $k({\vec \lambda};t)=\rho_s$, which immediately implies decoupling between 
$\eta({\vec R_{c}}+{\vec \lambda})$ and $P({\vec R_{c}},\eta;t)$ in
Eq.(\ref{defk}), then we  obtain from Eq.(\ref{vitesse}) a trivial mean-field-type result
\begin{equation}
V_{c}^{(0)}=(p_1-p_{-1})(1-\rho_s)\frac{\sigma}{\tau},
\label{vmf}
\end{equation}
which states that the frequency of jumps of the carrier particles ($\tau^{-1}$) only gets renormalized 
by a factor   $1-\rho_s$, which gives the fraction of successful jumps. 

The salient feature of our model is that there are essential backflow effects. The carrier effectively
perturbs the spatial distribution of the "environment" particles so that stationary density profiles
emerge. This can be contrasted to the  earlier dynamic
 percolation models 
\cite{drugerA,drugerB,zwanzig,sahimi,chatterjee,granekB,garcia,hilferA,granekS} in which the carrier had no impact on
the embedding medium and hence 
there was no re-arrengement of the host medium around the carrier particle. 
As a  consequence   $k({\vec \lambda};t) \neq \rho_s$,  and $k({\vec
\lambda};t)$  approaches
 $\rho_s$ only at infinite separations from the carrier, i.e. when 
$|{\vec \lambda}| \to\infty$.  Therefore, we rewrite Eq.(\ref{vitesse}) in the form
\begin{equation}
V_{c}(t)=V_{c}^{(0)}-\frac{\sigma}{\tau}\{p_1(k({\vec e_1};t)-\rho_s)-p_{-1}(\rho_s-k({\vec e_{-1};t}))\},
\label{gencorrneg}
\end{equation}
which shows
 explicitly the deviation of the  mean velocity of the carrier from the  mean-field-type result in 
Eq.(\ref{vmf}) due to the formation of the density profiles.

\subsection{Evolution equations of the pair correlation functions.}

From Eq.(\ref{vitesse}) it follows that in order to obtain $V_{c}(t)$, it suffices to compute
 $k({\vec e_{\pm1}};t)$. Consequently, we have to evaluate the equation governing the time evolution of the
pair correlation functions. Multiplying both sides of Eq.(\ref{eqmaitresse}) by 
 $\eta({\vec R_{c}})$ and summing over all configurations  $({\vec R_{c}},\eta)$, we find 
  that $k({\vec \lambda};t)$ obeys
\begin{eqnarray}
\partial_tk({\vec \lambda};t)&=& \frac{l}{6 \tau^*} \sum_{\mu}(\nabla_\mu-
\delta_{{\vec \lambda},{\vec e}_{\mu}}\nabla_{-\mu})k({\vec
\lambda};t)-\frac{(f+g)}{\tau^*}k({\vec \lambda};t)+\frac{f}{\tau^*}+\nonumber\\
&+&\frac{1}{\tau}\sum_{\mu} \sum_{{\vec R_{c}},\eta}p_\mu\left(
1-\eta({\vec R_{c}+e}_{ \mu})\right)\nabla_\mu\eta({\vec R_{c}}+{\vec \lambda})P({\vec R_{c}},\eta;t),
\label{evolk}
\end{eqnarray}
where  $\nabla_\mu$ denotes the ascending finite difference operator  of the form
\begin{equation}
\nabla_\mu f({\vec \lambda}) \equiv f({\vec \lambda}+{\vec e}_{ \mu})-f({\vec \lambda}),
\label{nabla}
\end{equation}
and
\begin{equation}
\delta_{{\vec r},{\vec r'}} = \left\{\begin{array}{ll}
1,     \mbox{ if the site ${\vec r}={\vec r'}$} \nonumber\\
0,     \mbox{ otherwise.}
\end{array}
\right.
\end{equation} 
The  Kroneker-delta term $\delta_{{\vec \lambda},{\vec e}_{\mu}}$ signifies 
that the 
evolution of the pair correlations, Eq.(\ref{evolk}), proceeds differently at large separations and at the
immediate vicinity of the carrier. This stems
 from the asymmetric hopping rules of the carrier particle defined by Eq.(\ref{trans}).
  
Note next that the contribution in the second line in Eq.(\ref{evolk}), which is associated with the 
biased diffusion of the carrier, appears to be non-linear with respect to the occupation numbers, such that the
pair correlation function gets effectively coupled to the evolution of the third-order correlations of the
form
\begin{equation}
T({\vec \lambda},{\vec e}_{\nu};t)\equiv\sum_{{\vec R_{c}},\eta}\eta({\vec R_{c}}+{\vec \lambda})\eta({\vec  R_{c}}
+{\vec e}_{\nu})P({\vec R_{c}},\eta;t).
\end{equation}
That is, Eq.(\ref{evolk}) is not closed with respect to the pair correlations but rather represents a
first equation in the infinite hierachy of coupled
equations for higher-order correlation functions. One faces, therefore, the problem of solving an infinite
hierarchy of coupled differential equations and needs to resort to an approximate closure scheme.

\subsection{Decoupling Approximation}

Here we resort to the simplest non-trivial closure approximation, based on the decoupling of the third-order
correlation functions into the product of pair correlations. More precisely, we assume that for 
 ${\vec \lambda}\neq{\vec e}_{\nu}$, the third-order correlation fulfils
\begin{eqnarray}
&&\sum_{{\vec R_{c}},\eta}\eta({\vec R_{c}}+{\vec \lambda})\eta({\vec R_{c}}+{\vec e}_{ \nu})P({\vec R_{c}},\eta;t)\nonumber\\
&\approx&\left(\sum_{{\vec R_{c}},\eta}\eta({\vec R_{c}}+{\vec \lambda})P({\vec R_{c}},\eta;t)\right)
\left(\sum_{{\vec R_{c}},\eta}\eta({\vec R_{c}}+{\vec e}_{\nu})P({\vec R_{c}},
\eta;t)\right),
\end{eqnarray}
or, in other words, 
\begin{equation}
\label{decouplage}
\sum_{{\vec R_{c}},\eta}\eta({\vec R_{c}}+{\vec \lambda})\eta({\vec R_{c}}+{\vec e}_{ \nu})P({\vec R_{c}},\eta;t)\nonumber\\
   \approx k({\vec \lambda};t)k({\vec e}_{\nu};t)
\end{equation}
The approximate closure 
in Eq.(\ref{decouplage})  has been already employed for studying related 
models of biased carrier diffusion in hard-core lattice gases and has been
shown to provide quite an
accurate description of both the dynamical and stationary-state behavior.
The decoupling in Eq.(\ref{decouplage})
 was first introduced in Ref.\cite{burlatsky}
 to determine the properties of a driven 
carrier diffusion in a
one-dimensional hard-core lattice gas with a conserved number of
particles, i.e.  without an exchange of particles with the reservoir.
Extensive numerical simulations performed 
in Ref.\cite{burlatsky} have demonstrated
that such a decoupling is quite a plausible
approximation for the model under study. 
Moreover,  rigorous
probabilistic analysis of Ref.\cite{olla} has shown 
that for this model the results
 based on the  decoupling scheme in Eq.(\ref{decouplage})
are exact. 
Furthermore, the same closure procedure
has been recently applied  to study spreading 
of a hard-core lattice gas from a
reservoir attached to one of the lattice sites \cite{spreading}. Again, a very good
agreement between the analytical results and the numerical data has been found.
Next, the decoupling in Eq.(\ref{decouplage}) has been used 
in a recent analysis of 
a biased carrier dynamics in
 a one-dimensional model of an adsorbed monolayer in contact
with a vapour phase \cite{benichou}, i.e. a one-dimensional version of the model
to be studied here. Also in this case an excellent agreement
has been observed between the analytical 
predictions  and the Monte Carlo simulations data \cite{benichou}. 
We now show  that the approximate closure of the hierarchy of the evolution equations
  in Eq.(\ref{decouplage}) allows us to reproduce
 in the limit $f,g=0$ and $f/g = const$
the results of Refs.\cite{nakazato} and \cite{elliott}, which are known (see e.g. Ref.\cite{kehr}) to provide a very
good approximation for the carrier diffusion coefficient in three-dimensional hard-core
lattice gases with arbitrary particle density.
 We expect therefore that such a closure scheme will render a plausible
description of the carrier dynamics in a 
three-dimensional generalized dynamic percolation 
model. We 
base our further analysis on this approximation.

Making use of Eq.(\ref{decouplage}), we  find  from Eq.(\ref{evolk})
that the pair correlations obey the following equations:
\begin{equation}
 \partial_tk({\vec \lambda};t)=\frac{l}{6 \tau^*} 
\tilde{L}k({\vec \lambda};t)+\frac{f}{\tau^*},
\label{systemek1}
\end{equation}
which hold for all lattice sites except for those at the immediate vicinity of the carrier, i.e. for all ${\vec \lambda}$
except for 
${\vec \lambda} = \{{\bf 0},{\vec e}_{\pm 1},{\vec e}_{\pm 2},{\vec e}_{\pm 3}\}$, while
 at the sites adjacent to the carrier one has
\begin{equation}
\partial_tk({\vec e}_{ \nu};t)=\frac{l}{6\tau^*} \Big(\tilde{L}+
A_\nu(t)\Big)k({\vec e}_{ \nu};t)+\frac{f}{\tau^*},
\label{systemek2}
\end{equation}
where $\nu=\{\pm1,\pm 2,\pm 3\}$. The operators $\tilde{L}$ and coefficients $A_\nu(t)$ 
are given explicitly by 
\begin{equation}
  \tilde{L}\equiv\sum_\mu A_\mu(t)\nabla_\mu-\frac{6(f+g)}{l},
  \end{equation}
and 
\begin{equation}
A_\mu(t) \equiv1+\frac{6\tau^*}{l\tau}p_\mu(1-k({\vec e}_{ \mu};t)),
\label{defA}
\end{equation}
where $\nabla_\mu$  has been defined previously in Eq.(\ref{nabla}), $\mu=\{\pm1,\pm 2,\pm 3\}$. It is important to emphasize that
all coefficients $A_\mu(t) = A_\mu(E,V_c;t)$, i.e. are functions of both the applied field and the carrier velocity.

Now, several comments about equations (\ref{systemek1}) and  (\ref{systemek2}) are in order.
First of all, let us note that Eq.(\ref{systemek2}) represents, from the mathematical point of view, 
the boundary conditions for the general evolution equation  (\ref{systemek1}), imposed on the sites in the
immediate vicinity of the carrier. Equations (\ref{systemek1}) and (\ref{systemek2}) have a different
form since in the immediate vicinity of the carrier its asymmetric hopping rules perturb essentially the
"environment" particles dynamics. Equations (\ref{systemek1}) and  (\ref{systemek2}) possess some
intrinsic symmetries and hence the number of independent parameters can be reduced. Namely, reversing the
field, i.e. 
changing 
$ E\to -E$,  leads to the mere replacement of
$k({\vec e_1};t)$ by $k({\vec e_{-1}};t)$ but 
does not affect  $k({\vec e}_{ \nu};t)$ with $\nu = \{\pm 2, \pm 3\}$, which implies that
  \begin{equation}
k({\vec e_1};t)(-E)=k({\vec e_{-1};t})(E),\;\;\mbox{and}\;\;k({\vec e}_{ 
\nu};t)(-E)=k({\vec e}_{\nu};t)(E) \;\; \mbox{for}\;\; \nu = \{\pm 2, \pm3\},
  \label{generalparite}
  \end{equation}
Besides, since  the transition probabilities in Eq.(\ref{trans}) obey
\begin{equation}
p_2=p_{-2}=p_3=p_{-3}
\end{equation}
one evidently has that
\begin{equation}
\label{s1}
k({\vec e_2};t)=k({\vec e_{-2}};t)=k({\vec e_3};t)=k({\vec e_{-3}};t),
\end{equation} 
and,  by symmetry, 
\begin{equation}
\label{s2}
A_2(t)=A_{-2}(t)=A_3(t)=A_{-3}(t)
\end{equation}
which somewhat simplifies equations (\ref{systemek1}) and  (\ref{systemek2}).
Lastly, we note that despite the fact that using the decoupling scheme in
Eq.(\ref{decouplage}) we effectively close the system of equations on the level of the pair
correlations, the solution of Eqs.(\ref{systemek1}) and (\ref{systemek2}) still poses serious technical 
difficulties. Namely, 
these equations are strongly non-linear with respect to the carrier velocity, which introduces the gradient term on
the rhs of the evolution equations for the pair correlation, and  depends by itself on the values of the
"environment" particles densities in the immediate vicinity of the carrier. Below we discuss a solution to this
non-linear problem, focusing on the limit $t \to \infty$. 
  
\section{Solution of the decoupled evolution equations in the stationary state.}
  
Consider the limit $t\to\infty$  and suppose that the density profiles and the stationary velocity of the carrier
have non-trivial stationary values
  \begin{equation}
  k({\vec \lambda})\equiv\lim_{t\to\infty}k({\vec \lambda};t), \;\;\;
  V_{c}\equiv\lim_{t\to\infty}V_{c}(t),\;\;\;\mbox{and}\;\;\; A_\mu = \lim_{t\to\infty}A_\mu(t).
  \end{equation}
Define next
the local deviations of $k({\vec \lambda})$ from the unperturbed density as
  \begin{equation}
  h({\vec \lambda})\equiv k({\vec \lambda})-\rho_s.
  \label{defh}
  \end{equation}
Choosing $h({\bf 0})=0$, we obtain  the following fundamental
system of equations: 
  \begin{equation}
\tilde{L}h({\vec \lambda})=0,
  \label{systemeh1}
  \end{equation}
which holds for  ${\vec \lambda} \neq \{{\bf 0},{\vec e}_{\pm 1},
{\vec e}_{\pm 2},{\vec e}_{\pm 3}\}$, 
while for the special sites adjacent to the carrier, i.e. for ${\vec \lambda} = \{{\bf 0},{\vec e}_{\pm 1},
{\vec e}_{\pm 2},{\vec e}_{\pm 3}\}$, one has
  \begin{equation}
 (\tilde{L}+A_\nu)h({\vec e}_{\nu})+\rho_s(A_\nu-A_{-\nu})=0,
  \label{systemeh2}
  \end{equation}
Equations (\ref{systemeh1}) and (\ref{systemeh2})
determine the spatial distribution of the deviation from the unperturbed density $\rho_s$ 
in the stationary state.  Note also that in virtue of the symmetry relations in Eqs.(\ref{s1}) and (\ref{s2}), 
$h({\vec e}_{\pm 2}) = h({\vec e}_{\pm 3})$
and $A_2 = A_{-2} = A_3 = A_{-3}$.

The method for solving the coupled non-linear 
Eqs.(\ref{vitesse}),(\ref{systemeh1}) and (\ref{systemeh2})
 is as follows:  We first solve these equations
supposing that the carrier
stationary velocity is a given parameter, or, in other words, 
assuming that $A_\nu$ entering Eqs.(\ref{systemeh1}) and (\ref{systemeh2}) are known.
In doing so, we obtain $h(\lambda)$ in the parametrized form
\begin{equation}
h({\vec \lambda})=h({\vec \lambda}; A_{\pm 1},A_{2}).
\label{hA}
\end{equation}
 Then, substituting into Eq.(\ref{hA}) particular values ${\vec \lambda}=\{{\vec e}_{\pm 1},{\vec e}_{\pm 2},{\vec e}_{\pm 3}\}$ 
and making use of the definition of $A_\mu$ in 
Eq.(\ref{defA}), we find a system of three linear equations with three
 unknowns of the form
  \begin{equation}
A_\nu=1+\frac{6\tau^*}{l\tau}p_\nu\Big(1-\rho_s-h({\vec e}_{\nu}; 
A_{\pm 1},A_{ 2})\Big),
  \end{equation}
where  $\nu=\{\pm1,2\}$, 
which will allow us to define all $A_\nu$ explicitly (and hence, all $h(\vec e_{\nu}))$. 
 Finally, substituting the results into Eq.(\ref{vitesse}), which can be written down in
terms of $A_\nu$ as
  \begin{equation}
  V_{c}=\frac{l\sigma}{6\tau^*}(A_1-A_{-1}),
  \label{vitimp}
  \end{equation}
we arrive at a closed-form equation determining implicitly 
the stationary velocity.

\subsection{Formal expression for the density profiles in the dynamic percolative environment 
as seen from the stationary moving carrier.}

The general solution of Eqs.(\ref{systemeh1}) and (\ref{systemeh2})
 can be most conveniently obtained by introducing the  generating
function 
  \begin{equation}
  H(w_1,w_2,w_3)\equiv\sum_{n_1,n_2,n_3}h(\vec{\lambda}) w_1^{n_1}w_2^{n_2} w_3^{n_3},
  \end{equation}
where $n_1$,$n_2$ and $n_3$ are the components of the vector ${\vec \lambda}$, ${\vec \lambda}
 = {\vec e}_1 n_1 + {\vec e}_2 n_2
 + {\vec e}_3 n_3$.  Multiplying both sides of Eqs.(\ref{systemeh1}) and (\ref{systemeh2})
 by $w_1^{n_1}w_2^{n_2}
w_3^{n_3}$ and performing summation, we find then that $H(w_1,w_2,w_3)$ is given explicitly by
  \begin{equation}
  H(w_1,w_2,w_3)=- l \frac{\sum_\nu\Big(A_\nu(w_{|\nu|}^{\nu/|\nu|}-1)h({\vec e}_{\nu})+
\rho_s(A_\nu-A_{-\nu})w_{|\nu|}^\nu\Big)}{l \sum_{\nu}A_\nu(w_{|\nu|}^{-\nu/|\nu|}-1)-6
(f+g)},
  \label{ddimH}
  \end{equation}
an expression which allows us to determine the stationary density profiles as seen from
 the carrier which moves with a constant
velocity $V_{c}$.

Inversion of the generating function defined by
 Eq.(\ref{ddimH}) yields then, after rather lenghty but straightforward calculations,  the following explicit result for the local deviation
from the unperturbed density:
  \begin{eqnarray}
\label{dev}
  h(\vec{\lambda})&=&\alpha^{-1}\Big\{\sum_\nu A_\nu h({\vec e}_{\nu})\nabla_{-\nu}
-\nonumber\\
&-&\rho_s (A_1-A_{-1})(\nabla_1-\nabla_{-1})\Big\} F(\vec{\lambda}),
  \end{eqnarray}
where $F(\vec{\lambda})$ is given by
  \begin{eqnarray}
  F(\vec{\lambda})=&&\left(\frac{A_{-1}}{A_1}\right)^{n_1/2}
\int_0^\infty e^{-x} {\rm I}_{n_1}\left(2 \frac{\sqrt{A_1 A_{-1}}}{\alpha}  
x\right) \times \nonumber\\ 
&\times&{\rm I}_{n_2}\left(2 \frac{A_2}{\alpha} x\right) 
{\rm I}_{n_3}\left(2 \frac{A_2}{\alpha} x\right)
{\rm d}x,
  \end{eqnarray}
and 
  \begin{eqnarray}
  \alpha&=&\sum_\nu A_\nu+\frac{ 6 (f+g)}{l} = \nonumber\\
&=& A_1 + A_{-1} + 4 A_2 + \frac{ 6 (f+g)}{l}
  \end{eqnarray}
Consequently, the particles density distribution as seen from the carrier moving with a constant velocity $V_{c}$
obeys
\begin{eqnarray}
\label{distr}
k(\vec{\lambda}) &=& \rho_s + \alpha^{-1}\Big\{\sum_\nu A_\nu h({\vec e}_{\nu}\nabla_{-\nu}
-\nonumber\\
&-&\rho_s (A_1-A_{-1})(\nabla_1-\nabla_{-1})\Big\} F(\vec{\lambda}),
\end{eqnarray}
where we have to determine three yet unknown parameters $A_1$, $A_{-1}$ and $A_2$.

To determine these parameters, we set in Eq.(\ref{dev}) 
${\vec  \lambda} = {\vec  e}_1$, ${\vec  \lambda} = {\vec  e}_{-1}$ and 
${\vec  \lambda} = {\vec  e}_{2}$, which results in the system of three closed-form equations determining the
unknown functions $A_{\nu}$, $\nu = \{\pm1,2\}$, 
\begin{equation}
\label{A}
 A_\nu=1+\frac{6\tau^*}{l\tau}p_\nu \left\{1-\rho_s
-\rho_s(A_1-A_{-1})\frac{\det\tilde{C}_\nu}{\det\tilde{C}}\right\},
\end{equation}
where $\tilde{C}$ is a square matrix of the third order defined as

\begin{equation}
\label{m}
\pmatrix{ A_1\nabla_{-1}F({\vec e}_{1}) - \alpha & A_{-1}\nabla_{1}F({\vec e}_{1}) &
                  A_{2}\nabla_{-2}F({\vec e}_{1}) \cr
        A_1\nabla_{-1}F({\vec e}_{-1}) & A_{-1}\nabla_{1}F({\vec e}_{-1}) - \alpha & A_{2}\nabla_{-2}F({\vec e}_{-1})\cr
         A_1\nabla_{-1}F({\vec e}_{2})  &A_{-1}\nabla_{1}F({\vec e}_{2}) &  
A_{2}\nabla_{-2}F({\vec e}_{2}) - \alpha}  
\end{equation}
while $\tilde{C}_\nu$  stands for the matrix obtained from 
$\tilde{C}$ by replacing the  $\nu$-th column by a column vector
$\left((\nabla_1-\nabla_{-1})F({\vec e}_{\nu})\right)_\nu$. Equation (\ref{distr}),
 together with the definition of the
coefficients $A_{\nu}$,  constitutes the first general result of our analysis defining the density
distribution in the percolative environment under study.

\subsection{General force-velocity relation.}

Substituting  Eqs.(\ref{dev}) and (\ref{m}) into (\ref{vitimp}), we find that
 the stationary velocity of the carrier particle is defined implicitly as the solution of equation:
\begin{equation}
\label{velo}
V_{c}=\frac{\sigma}{\tau}(p_1-p_{-1})(1-\rho_s)
\left\{1+\rho_s\frac{6\tau^*}{l\tau}\frac{p_1 \det\tilde{C}_1 - 
p_{-1} \det\tilde{C}_{-1}}{\det\tilde{C}}\right\}^{-1},
\end{equation}
where $\tilde{C}_{1}$ and $\tilde{C}_{-1}$ are the following square matrices of the third order:

\begin{equation}
\label{U}
\tilde{C}_{1} = \pmatrix{(\nabla_1-\nabla_{-1})F({\vec e}_{1})  & A_{-1}\nabla_{1}F({\vec e}_{1}) & A_{2}\nabla_{-2}F({\vec e}_{1}) \cr
      (\nabla_1-\nabla_{-1})F({\vec e}_{-1})   & A_{-1}\nabla_{1}F({\vec e}_{-1}) - \alpha & A_{2}\nabla_{-2}F({\vec e}_{-1})\cr
     (\nabla_1-\nabla_{-1})F({\vec e}_{2})    &A_{-1}\nabla_{1}F({\vec e}_{2}) &  
A_{2}\nabla_{-2}F({\vec e}_{2}) - \alpha}\\  
\end{equation}

and

\begin{equation}
\label{K}
\tilde{C}_{-1} = \pmatrix{ A_1\nabla_{-1}F({\vec e}_{1}) - \alpha &(\nabla_1-\nabla_{-1})F({\vec e}_{1})&
                  A_{2}\nabla_{-2}F({\vec e}_{1}) \cr
        A_1\nabla_{-1}F({\vec e}_{-1}) &(\nabla_1-\nabla_{-1})F({\vec e}_{-1})& A_{2}\nabla_{-2}F({\vec e}_{-1})\cr
         A_1\nabla_{-1}F({\vec e}_{2})  &(\nabla_1-\nabla_{-1})F({\vec e}_{2})&  
A_{2}\nabla_{-2}F({\vec e}_{2}) - \alpha}\\.  
\end{equation}
Equation (\ref{velo}) represents our second principal result defining the force-velocity relation in the dynamic
percolative environment for an arbitrary field and  arbitrary rates of the diffusive and renewal processes.

\section{Carrier velocity in the limit of small applied field $E$, friction coefficient and carrier diffusivity
 in dynamic percolative
environment. }

We consider now the case when the applied external field $E$ is small. Expanding the transition probabilities
$p_1$ and $p_{-1}$ in the Taylor series up to the first order in powers of the external field, i.e.
\be
p_{\pm 1} = \frac{1}{6} \pm \frac{\sigma \beta E}{12} +  {\mathcal O}\Big(E^2\Big), 
\ee
we find that $V_{c}$ defined by Eq.(\ref{vitimp}) follows
\be
V_{c} \sim \frac{\sigma}{6 \tau} \Big\{\sigma \beta E (1 - \rho_s) - (h(\vec{e}_1) - h(\vec{e}_{-1}))\Big\}.
\ee
On the other hand, Eq.(\ref{dev}) entails that 
\be
h(\vec{e}_1) - h(\vec{e}_{-1}) = \frac{2 \sigma \rho_s (1 - \rho_s) \tau^{*}}{l \tau \Big(\alpha_0 
{\cal L}(2 A_0/\alpha_0) - A_0\Big) + 2 \rho_s \tau} \beta E  +  {\mathcal O}\Big(E^2\Big),
\ee
where
\be
A_0 = lim_{E \to 0} A_\nu = 1 + \frac{\tau^{*}}{l \tau} (1 - \rho_s), 
\ee
and
\be
\alpha_0 = lim_{E \to 0}
\alpha = 6 \Big(1 + \frac{\tau^{*}(1 - \rho_s)}{l \tau} + \frac{f+g}{l}\Big), 
\ee
while 
\begin{eqnarray}
{\cal L}(x)&\equiv&\left\{\int_0^\infty e^{-t}{\rm I}_0^{2}(xt)\left({\rm I}_0(xt)-{\rm I}_2(xt)\right){\rm d}t\right\}^{-1}\nonumber\\
  &=&\left\{P({\bf 0};3x)-P(2{\vec e_1};3x)\right\}^{-1},
  \end{eqnarray}
$P({\vec r};\xi)$ being the generating function,
  \begin{equation}
  P({\vec r};\xi)\equiv\sum_{j=0}^{+\infty}P_j({\vec r})\xi^j,
  \end{equation}
of the probability $P_j({\vec r})$ that a walker starting at the origin and performing a Polya random walk on
the sites of a three-dimensional cubic lattice will arrive on the $j$-th step to
the site with the lattice vector  ${\vec r}$ \cite{3}.

Consequently, we find that in the limit  of a small applied field $ E$
the force-velocity relation in Eq.(\ref{velo}) attains the
physically meaningful form of the Stokes formula $E = \zeta V_c$, 
 which signifies that the frictional force exerted on the carrier
by the environment particles is  $\em viscous$.  
The effective friction coefficient
$\zeta$ is the sum of two terms, 
\begin{equation}
\zeta=\zeta_0 + \zeta_{coop}
\label{fricd}
\end{equation}
where the first term represents a mean-field-type result $\zeta_0
 = 6 \tau/\beta \sigma^2 (1 - \rho_s)$ (see Eq.(7)), while the second one, $\zeta_{coop}$, obeys
\begin{equation}
\zeta_{coop} =  \frac{12 \rho_s \tau^*}{\beta \sigma^2 l (1-\rho_s).
\Big(\alpha_0{\cal L}(2A_0/ \alpha_0)-A_0\Big)} 
\end{equation}
The second contribution has a more complicated origin and 
is associated with the cooperative behavior - formation of a inhomogeneous stationary particle distribution
around the carrier moving with constant velocity $V_{c}$. Needless to say, such an effect can not be observed within the framework of
previous models of dynamic percolation, since there the carrier does not influence the host medium dynamics 
\cite{drugerA,drugerB,zwanzig,sahimi,chatterjee,granekB,garcia,hilferA,granekS}.

Let us now compare the relative importance of two contributions, i.e. $\zeta_0$ and $\zeta_{coop}$, 
to the overall friction. In Fig.2 we
plot the ratio $\zeta/\zeta_0$ versus the creation rate $f$ for three different values of the density $\rho_s$, $\rho_s = 0.9, 0.7$ and $0.5$, while the
annihilation rate is prescribed by the relation $g = f (1 - \rho_s)/\rho_s$.
This figure
shows that the cooperative behavior clearly dominates at small and moderate $f$ (which entails also small values of $g$), 
while for larger $f$, when $\zeta/\zeta_0$ tends to $1$,
 the mean-field behavior
becomes most important. The cooperative behavior also appears to be more pronounced at larger densities $\rho_s$. 

Consider next some analytical estimates. We start with 
the situation, in which diffusion of the environment particles is suppressed, i.e. when $l = 0$. In this case,
we get
\begin{equation}
\frac{\zeta_{coop}}{\zeta_0} = \frac{2 \rho_s}{(1 - \rho_s)\Big(\frac{2}{y} {\cal L}(y) - 1\Big)},
\end{equation}  
where
\begin{equation}
y = \frac{1}{3} \Big(1 + \frac{\tau}{\tau^*}\frac{(f + g)}{(1 - \rho_s)}\Big)^{-1}.
\end{equation}
Suppose first that $\rho_s$ is small, $\rho_s \ll 1$. Then, $y \approx 1/3(1+\tau/\tau^*(f+g))$ and we can distinguish between
two situations: when $\tau  \ll (f+g)/\tau^*$, i.e. when the carrier moves faster than the environment re-organizes itself, and
and the opposite limit, $\tau  \gg (f+g)/\tau^*$, when the environment changes  very rapidly compared to the motion of 
the carrier. In the former case we find that $y \approx 1/3$, which yields $\zeta_{coop}/\zeta_0 \approx 2 \rho_s/(6 {\cal
L}(1/3) - 1)$, ${\cal
L}(1/3) \approx 0.7942$, while in the latter case we have $y \approx \tau^*/3 \tau (f +g)$ and 
$\zeta_{coop}/\zeta_0 \approx \rho_s \tau^*/3 \tau (f + g)$. Note, that in both cases the ratio $\zeta_{coop}/\zeta_0$ appears to
be small, which signifies that at small densities $\rho_s$ the mean-field friction dominates. Such a result is consistent with the behavior depicted
in Fig.2 and is not
counterintuitive, of course, since in the absence of the particles' diffusion, which couples effectively the density evolution at
different lattice sites, no significant cooperative behavior can emerge at small densities. On the other hand, 
at relatively high densities $\rho_s \sim 1$
and $\tau/(1 - \rho_s) \gg \tau^*/(f +g) \gg \tau$, when the carrier moves at much
faster rate than the host medium reorganizes itself, we find that $\zeta_{coop}/\zeta_0 \approx \tau^*/3 \tau (f + g) \gg 1$. 
This result stems from the circumstance that in sufficiently dense environments
modeled by dynamic percolation a highly inhomogeneous density profile emerges even in the absence of particles diffusion; Here, on the 
one
hand, the carrier perturbs significantly the particle density in its immediate vicinity. On the other hand, 
the density perturbance created by the carrier does not shift the global balance between creation and annihilation
events, i.e. the mean particle density still equals $\rho_s$, as we set out to show in what follows. The latter constraint
induces then appearence of essential correlations in particles distribution and hence, appearence of cooperative behavior.

Let us consider the opposite case when the renewal processes are not allowed, which means that 
the particles number is conserved and local density in the percolative environment evolves only due to particles diffusion. In
this case we find
\begin{equation}
\frac{\zeta_{coop}}{\zeta_0} = \frac{2 \tau^* \rho_s}{(l \tau + \tau^* (1- \rho_s)) \Big(6 {\cal L}(1/3) - 1\Big)}
\end{equation}
Here, the ratio  $\zeta_{coop}/\zeta_0$ can be large and the "cooperative" friction dominates the mean-field one
 when $l \tau \ll \tau^* (3 \rho_s -1)$, which happens, namely, at sufficiently high densities and in the limit when the carrier
moves at a much faster rate than the environment reorganizes itself. Otherwise, the mean-field friction prevails.

To estimate the carrier particle diffusion coefficient $D_c$ we assume
the validity of the Einstein relation, i.e. $\beta D_{c} = \zeta^{-1}$. 
We find that, in the general case,  the carrier diffusion coefficient $D_{c}$ reads
\begin{equation}
  D_{c}=\frac{\sigma^2(1-\rho_s)}{6\tau}\left\{1-\frac{2\rho_s\tau^*}{l\tau}\Big(\alpha_0{\cal
L}(2A_0/\alpha_0)-1+\frac{\tau^* (3\rho_s-1)}{l\tau}\Big)^{-1}\right\}
  \end{equation}
In the particular case of  conserved particles number, when $f,g \to 0$ but their ratio $f/g$ is kept fixed, $f/g =
\rho_s/(1-\rho_s)$, the latter equation reduces to the classical result
  \begin{equation}
\label{nk}
  D_{c}^{NK}=\frac{\sigma^2(1-\rho_s)}{6\tau}\left\{1-\frac{2\rho_s\tau^*}{l\tau} \Big(6 A_0{\cal
L}(1/3)-1+\frac{\tau^*  (3\rho_s-1)}{l\tau}\Big)^{-1}\right\},
  \end{equation}
obtained earlier in Refs.\cite{nakazato} and \cite{elliott} by different analytical techniques. The result in Eq.(\ref{nk}) is
known to be exact in the limits $\rho_s \ll 1$ and $\rho_s \sim 1$, and serves as a very good approximation for the
self-diffusion coefficient in hard-core lattice gases of arbitrary density \cite{kehr}.

It seems also interesting to analyse
 how random annihilation and creation of particles can modify the self-diffusion coefficient
compared
to the situation when the  particles number is conserved. In Figure 3 we plot the ratio $D^{NK}_c/D_c$ ($= \zeta/\zeta_{NK}$) versus the
creation rate $f$ for three different values of the density $\rho_s$, $\rho_s = 0.9, 0.7$ and $0.5$. Again, the value of the annihilation rate $g$
is prescribed by the relation $g = f (1 - \rho_s)/\rho_s$. Figure 3 shows that the
renewal processes affect considerably the friction coefficient and the ratio $\zeta/\zeta_{NK}$ 
deviates strongly from the unity with the growth of the creation rate. 
The overall friction also falls off when the density increases.

Finally,  in the
absence of particle diffusion (fluctuating-site percolation), 
our result for the carrier particle diffusion coefficient reduces to
\begin{equation}
  D_{c}^{per}=\frac{\sigma^2(1-\rho_s)}{6\tau}\left\{1-2\rho_s \Big(4[(1-\rho_s)+(f+g)\tau/\tau^*]{\cal
L}(y)+3\rho_s-1\Big)^{-1}\right\}
  \end{equation}
Note, however, that this result only applies when both $f$ and $g$ are larger than zero, such that the renewal processes
take place. In fact, the underlying 
decoupling
scheme is only plausible in this case. Similarly to the approximate theories in Refs.\cite{nakazato} and \cite{elliott}, our
approach predicts that in the absence of the renewal processes $D_c^{per}$ vanishes only when $\rho_s \to 1$, which is an
incorrect behavior.

\section{Asymptotic behavior of the density profiles at large distances in front of and past the carrier.}
  
The density profiles at large separations in front of and past the carrier can be 
readily deduced from the asymptotical behavior of the following generating function  
\begin{equation}
  N(w_1)\equiv\sum_{n_1=-\infty}^{+\infty}h(n_1,n_2=0,n_3=0) w_1^{n_1}.
  \end{equation}
Inversion of Eq.(\ref{ddimH}) with respect to the symmetric coordinates $n_2$ and $n_3$ 
yields then 
\begin{eqnarray}
&&N(w_1)= \frac{\Big(A_1h(\vec e_1)+\rho_s(A_1-A_{-1})\Big)\Big(w_1-1\Big) +
 \Big(A_{-1}h(\vec e_{-1})-\rho_s(A_1-A_{-1})\Big)\Big(w_1^{-1}-1\Big)  }{\alpha - A_1 w_1^{-1} - A_{-1} w_1} \times
\nonumber\\
&\times& \int_0^\infty \exp[- x ] {\rm I}_0^{2}(\frac{2 A_2 }{\alpha -A_1w_1^{-1}
-A_{-1}w_1}  x){\rm d}x + 
 \frac{4 A_2 h({\vec e}_2 )}{ \alpha - A_1 w_1^{-1} - A_{-1} w_1}  \times \nonumber  \\
&\times&  \int_0^\infty   \exp[-  x]  {\rm I}_0(  \frac{2 A_2 }{\alpha -A_1w_1^{-1}
-A_{-1}w_1}   x)
\Big({\rm I}_1( \frac{2 A_2 }{\alpha -A_1w_1^{-1}
-A_{-1}w_1} x)- \nonumber\\
&-& {\rm I}_0(\frac{2 A_2 }{\alpha -A_1w_1^{-1}
-A_{-1}w_1} x)\Big){\rm d}x
\end{eqnarray}
We notice now that $N(w_1)$ is a holomorphic function in the region ${\cal W}_1 < w_1 < {\cal W}_2$, where 
\begin{eqnarray}
{\cal W}_1 = \frac{\alpha - 4  A_2}{2 A_{-1}} - \sqrt{
\Big( \frac{\alpha - 4  A_2}{2 A_{-1}}\Big)^{2} - \frac{A_1}{A_{-1}} }
\end{eqnarray}
and
\begin{eqnarray}
{\cal W}_2 = \frac{\alpha - 4  A_2}{2 A_{-1}} + \sqrt{
\Big( \frac{\alpha - 4  A_2}{2 A_{-1}}\Big)^{2} - \frac{A_1}{A_{-1}} }
\end{eqnarray}
As a consequence, the asymptotic behavior of $h(n_1,n_2=0,n_3=0)$ in the limit $n_1 \to \infty$ (resp. $n_1 \to - \infty$)
is controlled by the behavior of $N(w_1)$  in the vicinity of $w_1 =  {\cal W}_2$ (resp.  $w_1 =  {\cal W}_1$) (see, for
example, the analysis of the generating function singularities  developed in Ref.\cite{flajolet}).

\subsubsection{Asymptotics of the density profiles at large separations in front of the carrier.}
  
Consider first the asymptotic behavior of the density distribution of the "environment" particles 
at large separations in front of the carrier.
Using the fact that 
\be
\int^\infty_0 exp\Big[ - x\Big] {\rm I}_0(y x)
\Big({\rm I}_1(x)-
{\rm I}_0(x)\Big){\rm d}x 
\ee 
is a regular function when $y \to 1/2$, while
\be
\int^\infty_0 exp\Big[ - x\Big] {\rm I}_0^2(y x) {\rm d}x \to \frac{1}{\pi} \ln\Big(\frac{1}{1 -2 y}\Big), 
\ee
we find that 
\begin{eqnarray}
\label{l}
 N(w_1) &\sim&_{w_1 \to {\cal W}_2} \Big[\frac{\Big(A_1h(\vec e_1)+\rho_s(A_1-A_{-1})\Big)\Big({\cal W}_2-1\Big)}{4
\pi A_2 } +
\nonumber\\
&+&\frac{\Big(A_{-1}h(\vec e_{-1})-\rho_s(A_1-A_{-1})\Big)\Big({\cal W}_2^{-1}-1\Big)}{4 \pi A_2}\Big]
\ln\Big({\cal W}_2-w_1\Big) 
\end{eqnarray}
Then, (cf, Flajolet et al., Ref.\cite{flajolet}), 
we obtain the following asymptotical result
\begin{equation}
h(n_1,0,0)\sim_{n_1 \to \infty }\frac{K^+}{n_1}e^{-n_1/ \lambda_+},
\end{equation}
where
\be
\lambda_+\equiv \ln^{-1}\left(\frac{\alpha/2-2A_2}{A_{-1}}+\sqrt{\left(\frac{\alpha/2-2A_2}{A_{-1}}\right)^2-\frac{A_1}{A_{-1}}}\right),
\ee
and
\begin{eqnarray}
K^+ &=& \Big[\frac{\Big(A_1h(\vec e_1)+\rho_s(A_1-A_{-1})\Big)\Big({\cal W}_2-1\Big)}{4
\pi A_2 } +
\nonumber\\
&+&\frac{\Big(A_{-1}h(\vec e_{-1})-\rho_s(A_1-A_{-1}\Big)\Big({\cal W}_2^{-1}-1\Big)}{4 \pi A_2}\Big] > 0,
\end{eqnarray}
which signifies that the density of the "environment" particles in front of the carrier is higher than the average value $\rho_s$ and approaches 
$\rho_s$ at large separations from the carrier as an exponential function of the distance.

\subsubsection{Asymptotics of the density profiles at large separations past the carrier.}
  
We consider next  the asymptotic behavior of the
 "environment" particles density profiles past the carrier particle, which turns
out to be very different depending on whether the  dynamics of
 the percolative environment obeys the strict conservation of
the "environment" particles number or not (the renewal processes are suppressed or allowed). The sketch of this behavior is
presented in Fig.4.

\paragraph{Non-conserved particles number.}

In the case when  partciles  may disappear and re-appear on the lattice, one has that the root ${\cal W}_1 < 1$.
We find then, following essentially the same
lines as in the previous subsection, that
\begin{eqnarray}
\label{k}
 N(w_1) &\sim&_{w_1 \to {\cal W}_1} \Big[\frac{\Big(A_1h(\vec e_1)+\rho_s(A_1-A_{-1})\Big)\Big({\cal W}_1-1\Big)}{4
\pi A_2 } +
\nonumber\\
&+&\frac{\Big(A_{-1}h(\vec e_{-1})-\rho_s(A_1-A_{-1})\Big)\Big({\cal W}_1^{-1}-1\Big)}{4 \pi A_2}\Big]
\ln\Big(\frac{1}{w_1 -{\cal W}_1}\Big).
\end{eqnarray}
Hence,
in the
non-conserved case  the approach to the unperturbed value $\rho_s$ is also exponential when  $n_1 \to -\infty$, and follows 
 \begin{equation}
  h_{n_1,0,0}\sim_{n_1 \to -\infty} \frac{K^-}{|n_1|}e^{-|n_1|/ \lambda_-},
\end{equation}
where
\be
\lambda_-\equiv\ln^{ -1}\left(\frac{\alpha/2-2A_2}{A_{-1}}-\sqrt{\left(\frac{\alpha/2-2A_2}{A_{-1}}\right)^2-\frac{A_1}{A_{-1}}}\right)
\ee
and
\begin{eqnarray}
K^- &=& \Big[\frac{\Big(A_1h(\vec e_1)+\rho_s(A_1-A_{-1})\Big)\Big({\cal W}_1-1\Big)}{4
\pi A_2 } +
\nonumber\\
&+&\frac{\Big(A_{-1}h(\vec e_{-1})-\rho_s(A_1-A_{-1})\Big)\Big({\cal W}_1^{-1}-1\Big)}{4 \pi A_2}\Big] < 0
\end{eqnarray}
which implies that the particles density past the carrier is lower than the average. 
Note that, in the general case,  $\lambda_+ < \lambda_-$, which
means that the depleted region past the carrier is more extended in space than
 the  traffic-jam-like region in front of the carrier. 
The density profiles are therefore asymmetric with respect to the origin, $n_1 = 0$.  Since creation 
of particles is favored (suppressed) in depleted (jammed)
regions, while annihilation
is suppressed (favored), one might expect that this will shift the overall
 density in the system, i.e. the average density of the "environment"
particles will differ from $\rho_s$.  
Interestingly, the overall deviation, i.e. the sum of local deviations over the
volume of the system,  of the density of the
"environment" particles from the average value $\rho_s$, appears to be equal exactly to zero,  
\be
  H(w_1=1,w_2=1,w_3=1) \equiv 0,
\ee
and hence, the driven carrier does not perturb the global balance between creation and annihilation of the "environment" particles. This is not,
however, an $\em a \; priori$ evident result in view of the asymmetry of the density profiles.

\paragraph{Conserved particles number.}
  
Finally, we turn to the analysis of the shape of the density profiles of the
 percolative environment past the carrier in the particular limit when the
host medium evolves only due to diffusion, while creation and annihilation of particles
are completely suppressed. In this case, in which the particles
number is explicitly conserved, one has that for arbitrary value of the field and particles' 
average density, the root ${\cal W}_1 \equiv 1$ and, 
consequently, the
form of the generating function is qualitatively different from that in Eqs.(\ref{l}) and (\ref{k}), 
\begin{eqnarray}
\label{y}
 N(w_1) &\sim&_{w_1 \to 1^+} \Big[\frac{\Big(A_1h(\vec e_1)-A_{-1}h(\vec
e_{-1})}{4
\pi A_2} + \nonumber\\
&+& \frac{2\rho_s(A_1-A_{-1})\Big)}{4
\pi A_2 }\Big] (w_1 -1)
\ln\Big(\frac{1}{w_1 -1}\Big).
\end{eqnarray}
Equation (\ref{y}) implies that in the limit when the particle number is conserved the large-$n_1$ 
asymptotic behavior of $h_{n_1,0,0}$
 is described by an algebraic function of $n_1$ with a logarithmic correction; that is, 
\begin{equation}
  h_{n_1,0,0}\sim\frac{K_- \ln(|n_1|)}{n_1^2},
  \end{equation}
where $K_-$ is an $n_1$-independent constant.  Remarkably, the power-law decay of correlations
 implies existence of a quasi-long-range order in the percolative environment past the carrier. 
 In the conserved case
the mixing of the three-dimensional percolative environment 
is not very efficient and there are considerable memory effects - 
the host medium  remembers
the passage of the carrier on  large space and time scales. 
  
\section{Conclusions}

To conclude,  we have studied  analytically 
the dynamics of a carrier driven by an external field ${\vec E}$  in a three-dimensional
 environment modeled by dynamic percolation on  cubic lattice partially filled with mobile, 
hard-core "environment" particles which can spontaneously disappear and
reappear (renewal processes) in the system
 with some prescribed rates. Our analytical approach has been based on
the master equation, describing the time
 evolution of the system, 
which has allowed us to evaluate a system of coupled
dynamical equations 
for the carrier
velocity and a 
hierarchy of correlation functions. 
To solve these coupled equations, we have invoked an approximate closure scheme
based on the decomposition of the
third-order correlation functions into a product of pairwise correlations, which has
been  
first 
introduced in Ref.\cite{burlatsky} for  a related
 model of a driven carrier dynamics in a one-dimensional lattice gas 
with conserved particles number. 
Within the framework of this approximation,
we have derived a system of coupled, discrete-space equations describing evolution 
of the density profiles of the  environment, as seen from the  moving
carrier, and its velocity $V_{c}$. We have shown  that 
 $V_{c}$ depends on  the density of the "environment" particles  in front of and past the carrier.
Both densities depend on the
 magnitude of the velocity, 
as well as on the rate of the renewal and diffusive processes. 
As a consequence of such a non-linear coupling,    
in the general case, (i.e. for an arbitrary driving field and arbitrary rates of renewal and diffusive
processes), 
$V_{c}$ has been found only implicitly, 
as the solution of a  non-linear 
equation relating its value to the system parameters. 
This equation, which defines the force-velocity relation for the dynamic percolation under study, 
simplifies considerably   
in the limit of small applied field ${\vec E}$. We find that in this limit it attains the physically meaningful form of the
Stokes formula, which implies, in particular, that the frictional force exerted on the carrier by the  environment 
modeled by dynamic percolation is $viscous$. In this limit, 
the carrier velocity and the friction coefficient are calculated explicitly. In addition, we determine the self-diffusion
coefficient of the carrier in the absence of the field and show that it reduces to the well-know result of
Refs.\cite{nakazato} and \cite{elliott} in the limit when the particles number is conserved. 
Further more, we have found that  the density profile  around the carrier becomes strongly
inhomogeneous: the local density of the "environment"
 particles in front of the carrier is higher than the 
average and approaches the average value as an exponential
 function of the distance from the carrier. 
On the other hand, past the carrier 
the local density is lower than the average, and depending on
whether the number of particles 
is explicitly conserved or not, the local density past the carrier
 may tend to the average value either as an exponential or even as an
 $\em algebraic$ function of the distance. The latter reveals 
especially strong memory effects and strong 
correlations between the particle distribution in the
environment and the carrier position.

\newpage

{\Large Figure Captions.}

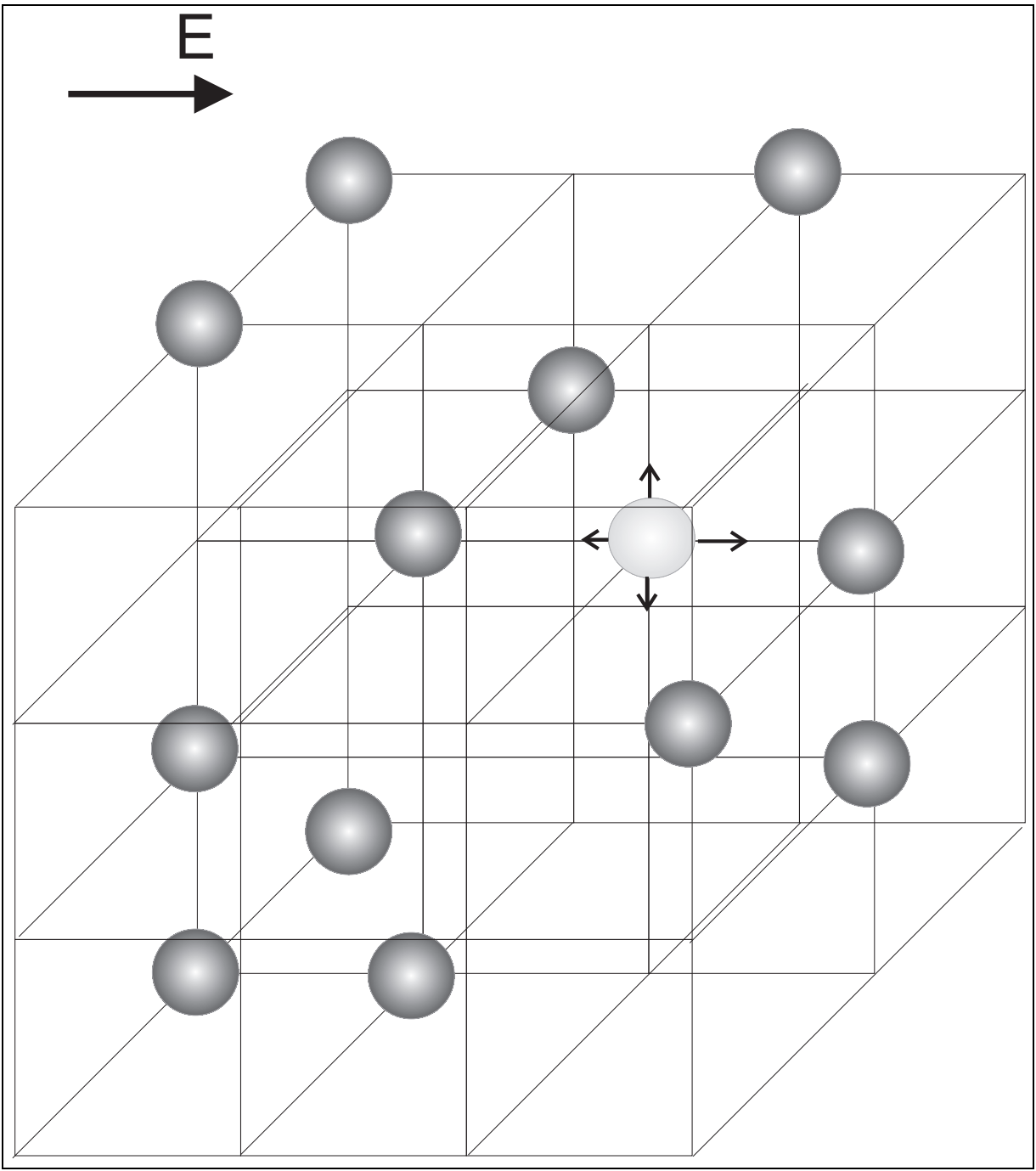
Fig.1. A generalized model of dynamic percolation. Grey spheres denote the hard-core 
"environment" particles, which perform symmetric random hopping among the sites of a simple cubic lattice, and can be
spontaneously annihilated and created. The lighter sphere is the carrier, which performs a biased  random walk
due to an external field ${\vec E}$, consrained by hard-core exclusion with the "environment" particles.

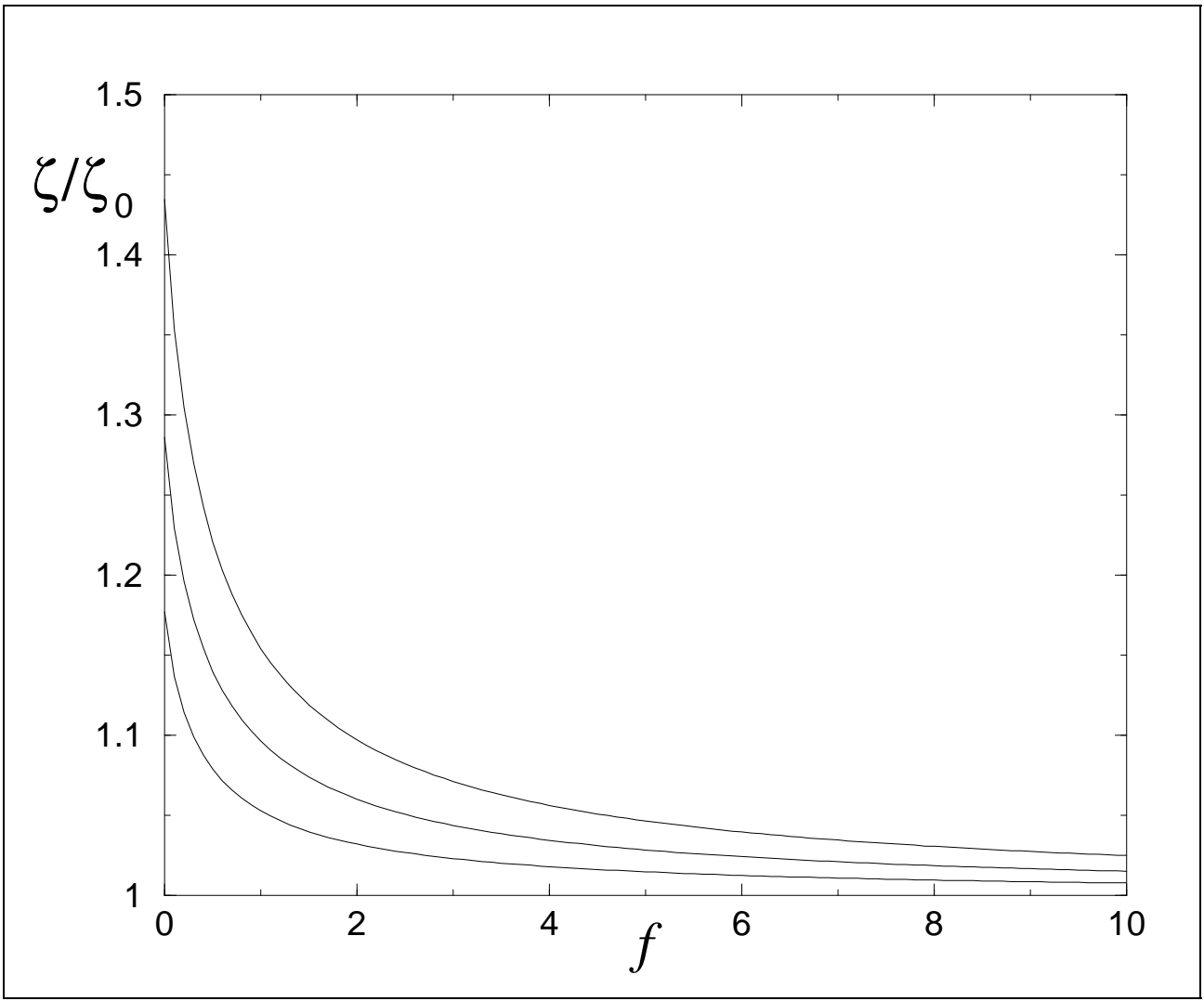
Fig.2. The ratio of the overall friction coefficient and the 
mean-field friction versus the creation rate
for three different values of the mean density $\rho_s$. The upper curve 
corresponds to $\rho_s = 0.9$, the
intermediate - to $\rho_s = 0.7$, and the lower - to $\rho_s = 0.5$.

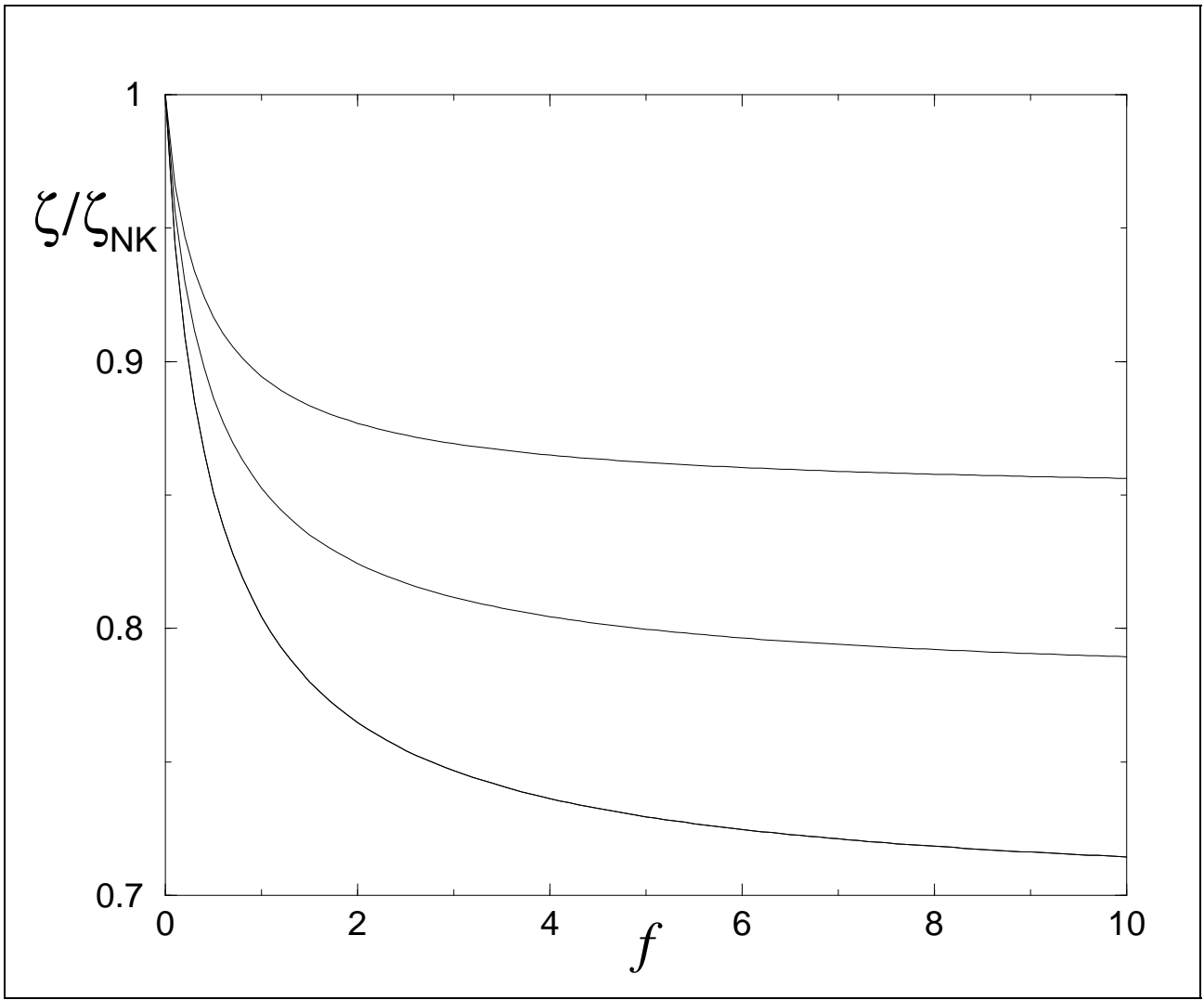
Fig.3. The ratio of the overall friction coefficient and 
the friction coefficient in the conserved particles number case versus the creation rate
for three different values of the mean density $\rho_s$. The upper curve corresponds to $\rho_s = 0.5$, the
intermediate - to $\rho_s = 0.7$, and the lower - to $\rho_s = 0.9$.

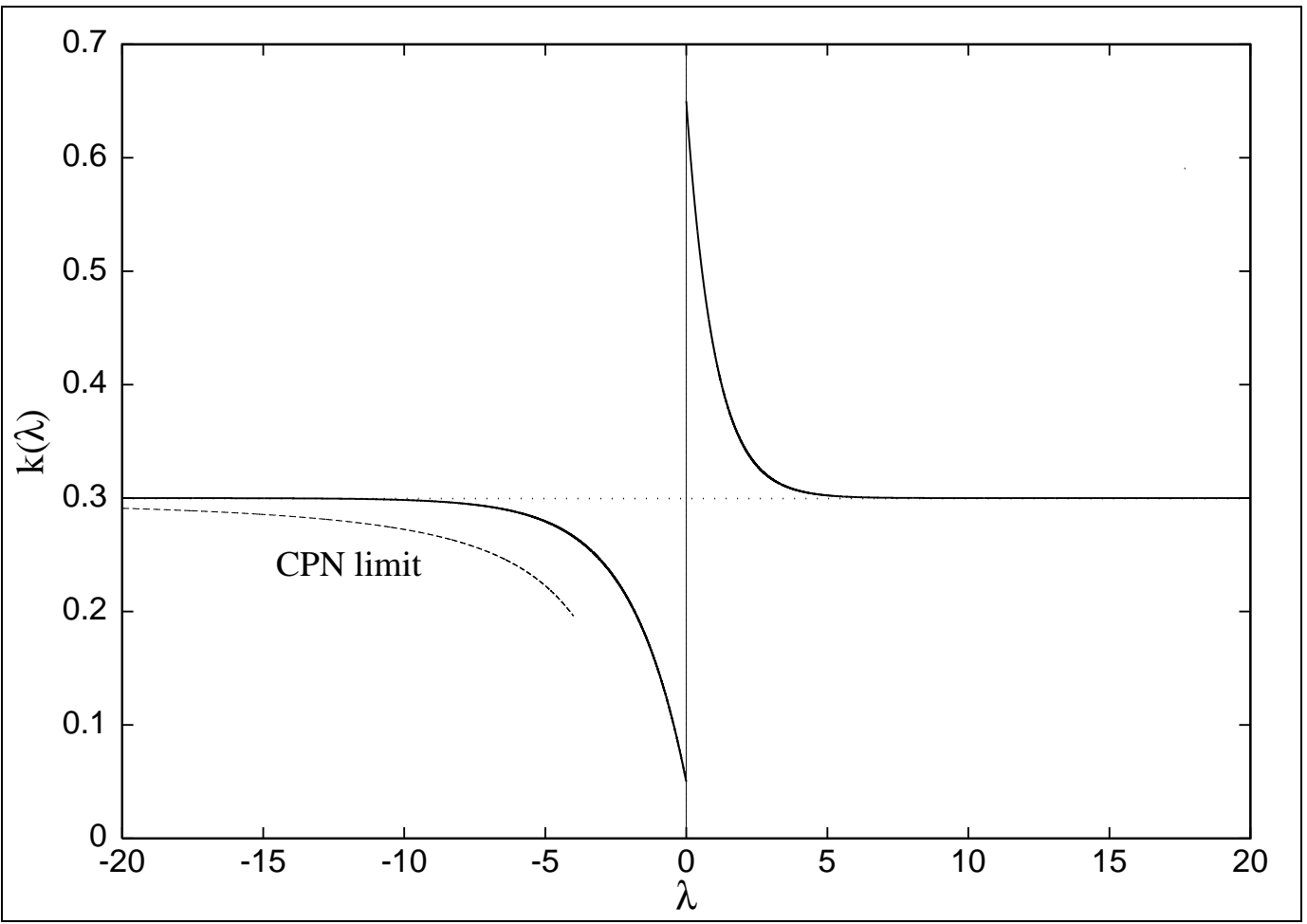
Fig.4. A sketch of the asymptotic density profiles in front of and past the stationary moving
carrier. The abbreviation CPN stands for the "conserved particles number". The two solid lines in the 
 $\lambda > 0$
and $\lambda < 0$ domains denote exponential profiles, Eqs.(63) and (67). The dashed line in the domain $\lambda < 0$ stands for the
algebraic law, Eq.(72).

\end{document}